\documentclass{PoS}
\usepackage{epsfig}
\PoS{PoS(LAT2005)083}
\title{Pseudoscalar Singlet Physics with Staggered Fermions }

\ShortTitle{Pseudoscalar singlet physics with staggered fermions } 

\author{\speaker{Eric B. Gregory}, Alan C. Irving, Craig McNeile, 
Steven Miller, and
Zbigniew Sroczynski\\
Theoretical Physics Division\\
Department of Mathematical Sciences\\
University of Liverpool\\
Liverpool,
L69 7ZL, UK\\
        E-mail: \email{egregory@liverpool.ac.uk}}

\abstract{We report on progress in measuring disconnected correlators 
associated with pseudoscalar flavor-singlet mesons. This will eventually 
allow us to compute the masses of the eta and eta' mesons. Flavor-singlet 
physics also presents an interesting test of the staggered fermion formulation,
as disconnected correlators are sensitive to whether the same action governs 
both sea quarks and valence quarks. It can also help test the 
validity of the ``fourth-root trick'' used in unquenched lattice calculations
where the number of flavors 
$N_f<4$.}

\FullConference{XXIIIrd International Symposium on Lattice Field Theory\\
		 25-30 July 2005\\
		 Trinity College, Dublin, Ireland}

\newcommand{\me}{\mathrm{e}}
\begin{document}
\section{Introduction}
Pseudoscalar flavor-singlet mesons are interesting for a number of reasons,
including the relation between the  $\eta'$ and $\pi$ mass difference 
and the U(1) axial anomaly \cite{Witten:1979vv,Veneziano:1979ec}, and 
the connection to the topological susceptibility. Unlike flavor non-singlet 
states, the propagator of flavor-singlet meson 
contains disconnected correlators, which are particularly challenging to 
measure precisely in lattice simulations. Hence flavor-singlet mesons present 
a rich proving ground for lattice simulations. To this end a number of 
simulations have been performed on the lattice using quenched Wilson 
\cite{Itoh:1987iy}, $N_f=2$ Wilson \cite{ McNeile:2000hf, Struckmann:2000bt,
Lesk:2002gd,Schilling:2004kg, Allton:2004qq}, 
and $N_f=2$ staggered fermion formulations
\cite{Venkataraman:1997xi,Kogut:1998rh}.

Simulations utilizing modern dynamical staggered fermion gauge configurations
are of particular interest. 
The library of $N_f=2+1$ staggered fermion gauge configurations 
from the MILC collaboration \cite{Bernard:2001av, Aubin:2004wf} contains
extremely light up and down quarks ($m_q/m_s<0.2$), simulated at fine
lattice spacings ($a\approx 0.11{\rm fm}$ and $a\approx 0.09{\rm fm}$).
These configurations, and other ensembles currently being generated provide
an unprecedented opportunity to measure mixed operators numerically determine 
the masses of both the 
$\eta$ and $\eta'$. Additionally, the sensitivity of disconnected correlators 
to sea quark loops offers a probe of the validity of the fourth-root trick 
in the staggered fermion formulation. 

With this in mind, we make a preliminary presentation of measurements of 
pseudoscalar flavor-singlet propagators on $N_f=2+1$-flavor improved staggered
fermion configurations. 
As the work is in preliminary stages, we concentrate 
on some  theoretical and technical issues in Sections 
\ref{theory} and \ref{meas}.
In Section \ref{results}  we describe our early results with some 
discussion of 
ongoing questions and issues, which we hope our continuing work will
resolve.

\section{Theoretical Considerations}
\label{theory}
The flavor-singlet pseudoscalar meson propagator
is expressed as 
\begin{equation}
G(x',x)=\langle\sum_{i}\overline{q}_i(x')(\gamma_5\otimes{\bf 1})q_i(x')\sum_{j}\overline{q}_j(x)(\gamma_5\otimes{\bf 1})q_j(x)\rangle,
\end{equation}
where the $(\gamma_5\otimes{\bf 1})$ denotes the meson as a pseudoscalar 
in spinor-space and a singlet in flavor-space. One can group the possible 
contractions into two classes:
$N_f$ connected terms with contractions connecting fields at $x$ and $x'$, 
and $N_f^2$ disconnected terms with the $q$s contracted with $\overline{q}$s
at the same space time points.
So the full propagator is:
\begin{equation}
\label{nflav_prop}
G = N_fC - N_f^2D,
\end{equation}
where $C$ and $D$ are the connected and disconnected correlators,
respectively. The minus sign is due to the additional fermion loop in 
$D$. The function $C$ is also the connected pion propagator with 
$(\gamma_5\otimes{\bf 1})$
operator. In full QCD, both $C$ and
$G$ will have leading behavior that decays exponentially:
\begin{equation}
C(t)\sim \me^{-m_\pi t} \qquad\mbox{ and  }\qquad G(t)\sim \me^{-m_{\eta'} t}.
\end{equation}
So the ratio of the disconnected to connected contributions to the singlet 
propagator behaves as
\begin{equation}
\label{DCratio}
R(t)\equiv\frac{N_f^2D(t)}{N_fC(t)}= \frac{N_fC(t)-G(t)}{N_fC(t)}
= 1-A\exp\left[-(m_{\eta'}-m_\pi)t\right]
\end{equation}
in full QCD \cite{Venkataraman:1997xi}. The correct behavior of the full 
propagator depends on the sea 
quarks having the same number of flavors and same masses as the valence quarks.
If the ${\rm det}^{1/4}$ trick introduces some unexpected
pathology into the action of the sea quarks, we might expect to detect a 
deviation from the form of equation \ref{DCratio}. 
For example, in the
limit of quenched QCD $R$ should be linear:
\begin{equation}
R(t)=A+Bt.
\end{equation}

In non-degenerate flavor simulations one should employ obvious generalizations
of equations \ref{nflav_prop} and \ref{DCratio}. For example,
when $N_f$ is split into $N_q+N_s$ flavors of light and strange quarks 
we have:
\begin{eqnarray}
G &\longrightarrow& \left(N_qC_{qq} + N_sC_{ss}\right) 
- \left(N_q^2D_{qq} +N_s^2D_{ss} + 2N_qN_sD_{qs}\right)\nonumber\\
R&\longrightarrow&\frac{N_q^2D_{qq} +N_s^2D_{ss} + 2N_qN_sD_{qs}}{N_qC_{qq} + N_sC_{ss}}.
\end{eqnarray}

With staggered fermions, one must rescale the disconnected correlators by 
$1/4$ with respect to the connected
correlators to take account the two loops of four staggered valence tastes
in the disconnected diagrams compared to the single valence loop in the 
connected diagrams\cite{Venkataraman:1997xi}. 
In the discussion below such rescaling is implicit in $D$.

\section{Simulation and Measurement}
\label{meas}
We have begun measurement of singlet propagators on the following gauge 
ensembles:
\begin{center}
\setlength{\tabcolsep}{1.5mm}
\begin{tabular}{|l|l|l|l|l|}
\hline
$N_f$& $\beta$ & $L^3\times T$ & $am$ & $N_{\rm configs}$\\
\hline
0   & 8.00 & $16^3\times 32$ & 0.02 & 104\\
2   & 7.20 & $16^3\times 32$ & 0.02 & 268 \\
2+1 & 6.76 & $20^3\times 64$ & 0.007, 0.05 & 422\\
2+1 & 6.76 & $20^3\times 64$ & 0.01, 0.05 & 644\\
\hline
\end{tabular}
\end{center}

The $16^3\times 32$ configurations are small test lattices generated 
locally, while the $20^3\times 64$ configurations are
part of the library of MILC ``coarse lattices'' \cite{Bernard:2001av}.

In principle there are two choices of flavor-singlet pseudoscalar meson
operator available. These are the $(\gamma_4\gamma_5\otimes\bf{1})$ and 
the $(\gamma_5\otimes\bf{1})$.
The former 
is a three-link operator, with the quark and antiquark sources set on opposite 
corners of the spatial cube, while the latter is a four-link
operator, with the quark and antiquark situated on opposite corners of the 
{\em hyper}cube. The 
$(\gamma_4\gamma_5\otimes\bf{1})$ has a parity partner, namely
the scalar $({\bf 1}\otimes\gamma_4\gamma_5)$, which contributes an 
oscillating exponential 
to the pseudoscalar propagator. The parity partner
of the $(\gamma_5\otimes\bf{1})$, however, has exotic quantum numbers 
$J^{PC}=0^{+-}$, and contributes nothing to the pseudoscalar propagator.
Hence, only the $(\gamma_5\otimes\bf{1})$ is well-suited for looking at 
quantities such as the ratio in equation \ref{DCratio}.
We formulate the operators in a gauge invariant way, using symmetric 
covariant shifts to displace the quark and antiquark sources to their 
respective positions on the hypercube.

We measure the connected diagrams using standard point sources. We measure 
the the disconnected diagrams with a stochastic volume source method
\cite{Farchioni:2004ej}. We define a
source field $\eta$ whose value at every lattice site is drawn from a gaussian
distribution. Then we solve for 
\begin{equation}
\mathcal{O}(t)= \frac{1}{N_{\rm src}}\sum_i^{N_{\rm src}}\sum_{x,x_4=t}\sum_{y,y_4=t}
{\rm Tr} \eta^{\dagger(i)}_y \Delta_{\gamma_r\otimes{\bf 1}} M^{-1}_{yx} \eta^{(i)}_x,
\end{equation}
where $M$ is the fermionic matrix and $\Delta_{\gamma_r\otimes{\bf 1}}$ is the 
staggered meson operator that effects the four-link shifts and the 
Kogut-Suskind phasing appropriate to the $\gamma_r\otimes{\bf 1}$ meson.
We average over a number of different source fields $\eta$, then compute
the disconnected correlator:
\begin{equation}
D(\Delta t) =  \langle \mathcal{O}(t) \mathcal{O}(t+\Delta t)\rangle
\end{equation}
 
\begin{figure}[ptbh!]
\begin{center}
\rotatebox{270}{\resizebox{3.0in}{!}{\includegraphics{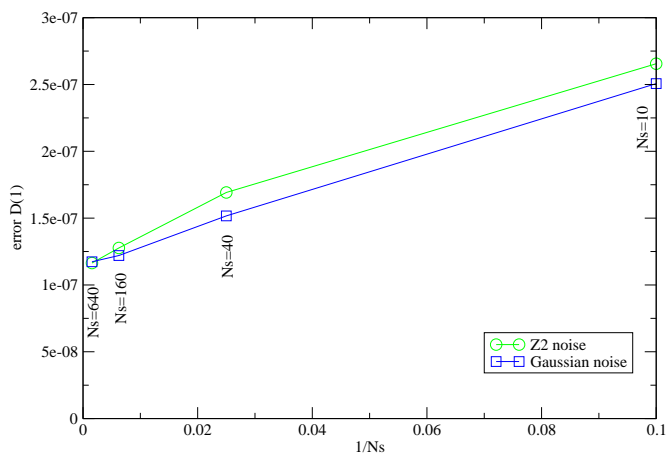}}}
\end{center}
\caption{\label{noise_errors} Error on the $D(t=1)$ for $N_f=2$ $\beta=7.2$
$am=0.20$ on $16^3\times 32$ lattices for gaussian and $Z_2$ noise as a function of the inverse number of sources.
}
\end{figure}

In tests on the small ($16^3\times 32$) lattices, we found that with as few
as 40 noise sources our error would be dominated by gauge noise even with 
several hundred configurations, so in subsequent runs we used 
$N_{\rm src}=40$. We also tested $Z_2$ noise and found that it produced 
noise errors that were systematically larger than that obtained with gaussian 
noise, as shown in Figure \ref{noise_errors}. Previously,
for Wilson fermions $Z_2$ noise was found to be better than Gaussian for the 
simplest stochastic estimators \cite{Dong:1993pk}.


\section{Results and Discussion}
\label{results}
Our results to date show clear signals for disconnected and connected 
correlators. 
For the 2+1-flavor ensembles, 
we formed the ratio:
\begin{equation}\label{21rat}
R(t)=\frac{4D_{qq}(t) + 4D_{qs}(t) + 1D_{ss}}{2C_{qq}(t) + C_{ss}(t)}.
\end{equation}
\begin{figure}[ptbh!]
\begin{center}
\rotatebox{270}{\resizebox{3.0in}{!}{\includegraphics{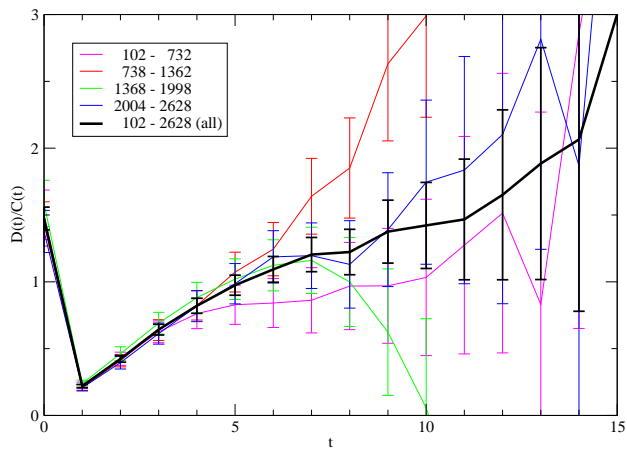}}}
\end{center}
\caption{ \label{ratio_plot}
$R(t)$ 
on $\beta=6.76$, $am=0.007,0.05$ on $20^3\times 64$ lattices for full 
time series 
(black), and four subsets (colored). Configurations indexed by trajectory 
number (6 trajectories per configuration).
}
\end{figure}
\begin{figure}[ptbh!]
\rotatebox{270}{\resizebox{2.0in}{6in}{\includegraphics{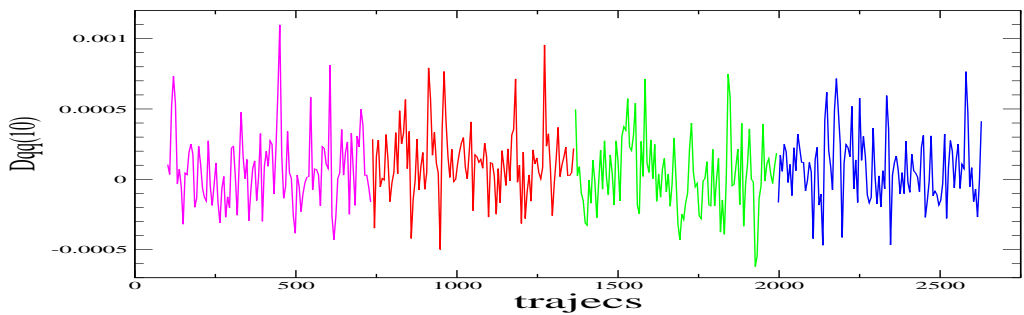}}}
\caption{\label{t_series} Time series for $D_{qq}(t=10)$ 
on $\beta=6.76$, $am=0.007,0.05$ on $20^3\times 64$ lattices. Same color scheme as previous figure. 
}
\end{figure}
An example of the resulting function is plotted as the dark curve in 
Figure \ref{ratio_plot}. Results for both $2+1$-flavor ensembles were 
qualitatively similar.

It is clear that our correlators as formulated 
do not form the components of a singlet  propagator which is positive at all 
$t$, as the magnitude of the disconnected correlators exceeds
that of the connected correlators for $t>5$. Correctly normalized 
correlators showing 
this behavior might illustrate a flaw in the simulated staggered sea. We have 
suspicions that the normalizations of our disconnected operators may 
instead be incorrect, however. As we compute the disconnected and connected 
correlators by different methods, we must account for different numerical 
factors in each, and reconciling them is non-trivial.  Further tests are 
being made.

Further uncertainty arises from long autocorrelations in the disconnected 
correlators. 
Figure \ref{ratio_plot} also shows the ratio plotted with four different 
subsets of the ensemble, each composed of a quarter of the time series. The 
large difference between bins suggests that one needs far more
gauge configurations to make a statement about the behavior of the $D$ to $C$
ratio. The autocorrelations are not visually apparent from inspection of 
the time series of the disconnected correlators, e.g. Figure \ref{t_series}, 
but spikes in otherwise small fluctuations seem to be enough 
to strongly affect $D/C$. It is also possible that with a longer time series
$R(t)$ may settle at a smaller asymptote than is apparent now.

With these uncertainties, we do not conclude that the pseudoscalar singlet 
correlators produced thus far cast doubt on the staggered formulation, 
nor are we at this
stage able to make statements about the lattice masses of the $\eta$ and 
$\eta'$.  It is evident that we will first have to confirm the correct 
normalizations, and then make measurements on extremely long time series
ensembles. We are in the process of generating $2+1$-flavor ensembles of 
$\sim 10^4$ configurations on the UKQCD's QCDOC machine. 
Additionally, we are investigating different optimization strategies 
\cite{Venkataraman:1997xi, Foley:2005ac} such as 
dilution  of the stochastic sources which may decrease 
disconnected correlator noise.

\end{document}